\documentclass[12pt,fleqn]{article}
\usepackage{latexsym}

\begin{document}
\title{Bell's Theorem - Why Inequalities, Correlations?}
\author{Noam Erez$\footnote{E-mail:nerez@physics.tamu.edu}$
 {\ }\\
\small  {\em \small Institute for Quantum Studies and
Department of Physics, Texas A\&M University,}\\ \small {College
Station, \small TX 77843-4242, USA} 
}

\maketitle
\begin{abstract} It is shown that Bell's counterfactuals admit joint
quasiprobability distributions (i.e. joint distributions exist, but may not be non-negative). 
A necessary and sufficient condition for the existence among them of
a true probability distribution (i.e. non-negative) is Bell's inequalities. This, in turn, is a 
necessary condition for the existence of local hidden variables. 
The treatment is amenable to generalization to examples 
of 'nonlocality without inequalities'.
\end{abstract}

\vspace{.1 in}

\section{Introduction}

Bell's derivation of his famous inequalities \cite{Bell1} of forty years ago
 hardly leaves room for improvement in terms of conciseness or elegance 
(see, however \cite{Peres} for a particularly clear derivation and discussion).
It is the purpose of this paper to give a straightforward ('brute force') 
derivation of them starting from very simple assumptions. The inequalities automatically follow 
as a necessary and sufficient condition for the impossibility of Bell type 
local realism, for the situation he envisaged\cite{Disclaimer}. 

To demonstrate the approach, let us first consider a toy problem. 
Let us assume we have two spin-1/2 particles in an EPR-Bohm\cite{EPR,Bohm} 
state, i.e. their spins are in a singlet state, and their position state 
corresponds to one of of them being localized in the vicinity of an observer named Alice, and the other
in the vicinity of a remote observer named Bob. Let us now assume that Alice has chosen a 
particular measurement to perform that can return either of two possible 
outcomes. Similarly Bob has chosen another such binary measurement. For 
convenience, let's label the two outcomes of each measurement as $\pm 1$.
Suppose further that Alice's outcomes have probabilities
\[ P[A=+1]=p^A_+,\ P[A=-1]=p^A_- \]where $p^A_{\pm}$ are determined by some theory 
(in particular we would be interested by outcomes predicted by Quantum 
Mechanics, but that is immaterial at this point). Bob's outcomes have 
probabilities $p^B_{\pm}$. We now pose the question: can a joint probability,
$\{ P^{AB}_{ab}|a,b\in \{+1,-1\} \} $ be defined such that the marginal 
probabilities it generates for $A,B$ coincide with $P^A$ and $P^B$.
In other words, $P^{AB}$ has to satisfy:

\begin{eqnarray}
P^{AB}_{++}+P^{AB}_{+-}&=&P^A_+\nonumber \\
(P^{AB}_{-+}+P^{AB}_{--}&=&P^A_-)\nonumber \\
P^{AB}_{++}+P^{AB}_{-+}&=&P^B_+\nonumber \\
(P^{AB}_{-+}+P^{AB}_{--}&=&P^B_-)\nonumber\\
P^{AB}_{++}+P^{AB}_{+-}+P^{AB}_{-+}+P^{AB}_{--} &= &1 \label{toyeq} 
\end{eqnarray}and 
\begin{equation} P^{AB}_{ab}\geq 0,\ a,b\in\{+1,-1\} \label{toyineq} 
\end{equation}The equations in 
parentheses are easily seen to be redundant - they are implied by the rest, and by the fact
that the marginals, being distributions, sum to 1.

We can answer in the affirmative immediately, since the 'product probability'
$P^{AB}_{ab} \equiv P^A_a P^B_b$ is indeed always a well defined probability, and 
has the desired marginals. This also implies that $P^{AB}$ is not, in general,
uniquely defined, since we could start by choosing a manifestly non-product
joint distribution and it would be different than the product distribution 
defined by its marginals. However, the point is, that Eqs.(\ref{toyeq}) can
be solved directly. These solutions can be called quasiprobabilities, since 
they are not necessarily non-negative. Inequalities (\ref{toyineq}) are just
the statement that the distribution correspond to a probability (i.e. be
non-negative). Since the constraints on $P^{AB}$ can always be satisfied,
there is no dependence on any particular assumptions about the marginal 
distributions. No matter what Quantum Mechanics predicts, it can be mimicked
by a local realistic model (see next section).

We shall see below that in Bell's scenario - the same as the toy problem, but
now Alice can choose to measure either $A_1$ or $B_1$, and Bob either
$B_2$ or $C_2$. The variables $B_1$ and $B_2$ will be chosen in a special
way, and now the predictions of Quantum Mechanics will be important.
While Quantum Mechanics does not define a joint probability 
$P^{A_1B_1B_2C_2}_{abcd}$ ($a,b,c,d\in\{\pm 1\} $), it does predict the 
joint distributions $P^{A_1B_2}, P^{A_1C_2}, P^{B_1C_2}$ and 
$P^{B_1B_2}$ ($B_1,2$ will be defined such that $P^{B_1B_2}_{ab}$ will be
nonzero only for $a=-b$). It will be seen that while the equations for $P^{A_1B_1B_2C_2}$
generalizing \ref{toyeq} can still be satisfied, the inequalities generalizing \ref{toyineq}
can only be satisfied for some marginal distributions. In particular, for some choices of
variables $A,B,C$, those constraints will clash with the predictions of Quantum Mechanics.
For binary valued pairs of variables, the probability distributions can be stated in terms of the
expectation values of the products: $\langle AB\rangle \equiv E(A_1 B_2)$, etc. (see below).

When expressed in this way, the constraints become identical to Bell's famous inequalities.
For n-valued measurables ($n>2$) or more than two particles, linear correlations would no longer suffice.
The fact that any marginal single observable distributions are compatible with a joint probability 
distribution remains true when we increase the number of such observables, and the number of values
they can take, as long as both remain finite. The simple construction of a joint product distribution
carries over to this case. Note that the single observables can be composite (e.g. vectors) but the 
different observables should be defined independently of each other (unlike $(A,B_2)$ and $(B_1,C)$, 
$B_1=-B_2$ used by Bell).

When we go to the continuous case, this no longer holds. As shown in\cite{Bertrand},
one way to define the Wigner quasiprobability distribution for the state of a particle with a one 
dimensional continuous degree of freedom, is simply to require that it generate the correct marginal 
distributions for all variables of the form $x_{\theta}\equiv \cos \theta x + \sin \theta p$ 
(for a very good exposition see \cite{Ulf}). 
These conditions define Wigner's distribution $W(q,p)$ uniquely. In general $W$ takes both positive
and negative values. Coherent states, which are arguably the closest to being ``classical'', are a notable
exception. ``Scr\"{o}dinger cat'' states display strong oscillations with notable negative dips. 
The analogy with what follows seems to be more than coincidental.

Following Bell's seminal paper, it was shown that with three particles in a particular entangled state
(GHZ state \cite{GHZ} and for a particular choice of observables, one can get a contradiction 
with the assumption of existence a local realist joint probability, not involving
inequalities. Instead, the existence of a local realist theory implies the existence of single events
that violate the predictions of QM. More recently, Hardy has shown\cite{Hardy} that one could
get similar results even with two spin-1/2 particles. In his example, local realism implies that either
some events must exist that violate the predictions of quantum theory, or other events (outcomes of
measurements) should never occur. Thus these tests involve no inequalities, besides perhaps whether 
some probability be larger than 0. Surprisingly, Hardy's construction works for almost all entangled 
states, the only exceptions being the maximally entangled ones! Those are precisely the states that
display maximal violation of Bell's inequalities. 

The direct derivation of Bell's inequalities provided here, shows that for maximally entangled states, 
no scheme involving (essentially) 3 observables (as in Bell's original derivation) can do
better than give precisely Bell's statistical inequalities as the local realist predictions. 
It might be hoped that a general analysis of the 4 observable case along these lines 
could shed some light on this intriguing complementarity between Bell's and Hardy's examples.

\section{The Condition for Existence of Quasi-Probabilities}

Let us assume a hidden variable $\lambda$ exists, that determines the outcome of the measurement of 
$A=\overrightarrow{\alpha} \cdot \overrightarrow{\sigma_1}$: $A=f(\overrightarrow{\alpha},\lambda)$, 
and similarly for $B=\overrightarrow{\beta} \cdot \overrightarrow{\sigma_i}$ ($i\in \{1,2\}$) and
$C =\overrightarrow{\gamma} \cdot \overrightarrow{\sigma_2}$. 
In other words, $A_1$,$B_i$ and $C_2$ are random variables in the same space.
The statistics of $\lambda$ determine a well defined probability distribution 
\mbox{$ P(A_1=a,B_2=b,C_2=c)$ $(a,b,c \in \{-1,1\})$}. We have suppressed $B_1$, because it will be assumed 
that, with probability 1, $B_2=-B_1$.
In what follows, we will use the notation 
$P^{ABC}_{+++} = P(A_1=+1,B_2=+1,C_2=+1)$, etc.
While it is assumed that this common distribution exists, it is also assumed that only two of the variables 
are simultaneously experimentally accessible. We note that $P^{ABC}$ determines the marginal distributions:
$P^{AB},P^{AC}$ and $P^{BC} \equiv P^{B_1C}=P^{(-B_2)C} $ ($P^{AB}_{ab}=\sum_{c=+,-}P^{ABC}_{abc}$,...).

The marginal probabilities are given by quantum theory (and verified experimentally).
The problem is to find $\{P^{ABC}_{abc} \}_{a,b,c=+,-}$ satisfying:

\begin{equation}
\left( \matrix{
   1 & 0 & 0 & 0 & 1 & 0 & 0 & 0 \cr 0 & 1 & 0 & 0 & 0 & 1 & 0 & 0 \cr 0 & 0 & 1 & 0 & 0 & 0 & 1 & 0 \cr 1 & 0 & 1 & 0 & 0 & 0 & 
   0 & 0 \cr 0 & 1 & 0 & 1 & 0 & 0 & 0 & 0 \cr 0 & 0 & 0 & 0 & 1 & 0 & 1 & 0 \cr 1 & 1 & 0 & 0 & 0 & 0 & 0 & 0 \cr 0 & 0 & 1 & 
   1 & 0 & 0 & 0 & 0 \cr 0 & 0 & 0 & 0 & 1 & 1 & 0 & 0 \cr 1 & 1 & 1 & 1 & 1 & 1 & 1 & 1 \cr  }\right)
\left( \matrix{P^{ABC}_{+++} \cr P^{ABC}_{++-} \cr P^{ABC}_{+-+} \cr P^{ABC}_{+--} \cr P^{ABC}_{-++} \cr
P^{ABC}_{-+-} \cr P^{ABC}_{--+} \cr P^{ABC}_{---} } \right) = 
\left( \matrix{P^{BC}_{++} \cr P^{BC}_{+-} \cr P^{BC}_{-+} \cr
P^{AC}_{++} \cr P^{AC}_{+-} \cr P^{AC}_{-+} \cr 
P^{AB}_{++} \cr P^{AB}_{+-} \cr P^{AB}_{-+} \cr 1}
\right)
\end{equation}

Which would make it a quasiprobability distribution. Note that $P^{BC}_{--},P^{AB}_{--},P^{AC}_{--}$ do not 
appear on the right hand side. That is because the equations for those lines are automatically satisfied 
when the others are, by the normalization of the marginal probabilities  (i.e. those equations were removed 
because they depend linearly on the rest). Referring this matrix equation as $M x= p$, we note that
$M$'s rank is only 7, so that if a solution exists, it is not unique. In other words, the
homogeneous equation $M x = 0$ has a unique solution (up to a multiplicative constant), 
$x_h=(-1,1,1,-1,1,-1,-1,1)$.
The condition for the existence of solutions is that the vector $p$ be in the column-space of $M$, or 
equivalently, orthogonal to its orthogonal complement. This orthogonal complement is spanned by the vectors
\[\{(-1, -1, 0, 0, 0, 0, 1, 0, 1, 0), (0, 0, 0, -1, -1, 0, 1, 1, 0, 0), 
(-1, 0, -1, 1, 0, 1, 0, 0, 0, 0)\}
\]The condition is therefore that the marginal probabilities satisfy the equations:

\begin{eqnarray}
P^{BC}_{++}+P^{BC}_{+-} & = & P^{AB}_{++}+P^{AB}_{-+} \nonumber \label{consistency} \\
P^{AC}_{++}+P^{AC}_{+-} & = & P^{AB}_{++}+P^{AB}_{+-} \nonumber \\
P^{BC}_{++}+P^{BC}_{-+} & = & P^{AC}_{++}+P^{AC}_{-+} 
\end{eqnarray}
These are simple consistency requirements. For example, the first equation follows from the requirement 
that both sides be equal to $P^{ABC}_{+++}+P^{ABC}_{-++}+P^{ABC}_{++-}+P^{ABC}_{-+-}$.
It will be seen in the next section that, remarkably, these conditions are satisfied by Bell's 
counterfactuals for a singlet state.

When these equations are satisfied, the family of quasiprobabilities, $x$, is given by 
$x=M^+ p+c x_h$, where $M^+$ is the pseudoinverse of $M$, and $x_h$ is the same as above, and $c$ any 
real number. 

The expression for $M^+$ is:
\begin{equation}
\left(
\matrix{ \frac{1}{4} & -\left( \frac{1}{8} \right)  & -\left( \frac{1}{8} \right)  & \frac{1}{4} & -\left( \frac{1}{8} \right)
      & -\left( \frac{1}{8} \right)  & \frac{1}{4} & -\left( \frac{1}{8} \right)  & -\left( \frac{1}{8} \right)  & \frac{1}
   {8} \cr -\left( \frac{1}{20} \right)  & \frac{13}{40} & \frac{1}{8} & -\left( \frac{1}{20} \right)  & \frac{13}{40} & \frac{1}
   {8} & \frac{7}{20} & -\left( \frac{3}{40} \right)  & -\left( \frac{3}{40} \right)  & -\left( \frac{1}{8} \right)  \cr -\left(
     \frac{1}{20} \right)  & \frac{1}{8} & \frac{13}{40} & \frac{7}{20} & -\left( \frac{3}{40} \right)  & -\left( \frac{3}
     {40} \right)  & -\left( \frac{1}{20} \right)  & \frac{13}{40} & \frac{1}{8} & -\left( \frac{1}{8} \right)  \cr \frac{1}
   {20} & -\left( \frac{9}{40} \right)  & -\left( \frac{9}{40} \right)  & -\left( \frac{3}{20} \right)  & \frac{3}{8} & -\left(
     \frac{1}{40} \right)  & -\left( \frac{3}{20} \right)  & \frac{3}{8} & -\left( \frac{1}{40} \right)  & \frac{1}{8} \cr 
    \frac{7}{20} & -\left( \frac{3}{40} \right)  & -\left( \frac{3}{40} \right)  & -\left( \frac{1}{20} \right)  & \frac{1}
   {8} & \frac{13}{40} & -\left( \frac{1}{20} \right)  & \frac{1}{8} & \frac{13}{40} & -\left( \frac{1}{8} \right)  \cr -\left(
     \frac{3}{20} \right)  & \frac{3}{8} & -\left( \frac{1}{40} \right)  & \frac{1}{20} & -\left( \frac{9}{40} \right)  & -\left(
     \frac{9}{40} \right)  & -\left( \frac{3}{20} \right)  & -\left( \frac{1}{40} \right)  & \frac{3}{8} & \frac{1}{8} \cr -\left(
     \frac{3}{20} \right)  & -\left( \frac{1}{40} \right)  & \frac{3}{8} & -\left( \frac{3}{20} \right)  & -\left( \frac{1}
     {40} \right)  & \frac{3}{8} & \frac{1}{20} & -\left( \frac{9}{40} \right)  & -\left( \frac{9}{40} \right)  & \frac{1}
   {8} \cr -\left( \frac{1}{4} \right)  & -\left( \frac{3}{8} \right)  & -\left( \frac{3}{8} \right)  & -\left( \frac{1}{4}
     \right)  & -\left( \frac{3}{8} \right)  & -\left( \frac{3}{8} \right)  & -\left( \frac{1}{4} \right)  & -\left( \frac{3}
     {8} \right)  & -\left( \frac{3}{8} \right)  & \frac{7}{8} \cr  }
\right)
\end{equation}
Due to the symmetries in Bell's problem, the equation will look much simpler for that case.

Finally, a probability distribution also has to satisfy the 8 \emph{inequalities}: 
$P^{ABC}_{abc}= (M^+ p+c x_h)_{abc} \geq 0$ ($a,b,c=+,-$). 
In the next section it will be shown that, for Bell's problem, these are equivalent to Bell's inequalities.

\section{Bell's Counterfactuals: The Singlet State}

Let our two-spin-$\frac{1}{2}$ system be in the singlet state $|\psi_{-}\rangle$, and the observables
$A,B,C$ be defined as above. Then the two-observable common distribution functions are equal to:

\begin{equation}
P^{A_1 B_2}_{ab}=\langle \pi^{A_1}_a \pi^{B_2}_b \rangle_{\psi_{-}} (a,b=+,-)
\end{equation}where $\pi^{A_1}_a$ is the projection operator 
$\frac{1 + a \overrightarrow{\alpha} \cdot  \overrightarrow{\sigma} }{2}$, etc.
So,
\begin{eqnarray}
P^{A_1 B_2}_{ab} & = & \frac{1}{4}\langle 1+ a\overrightarrow{\alpha} \cdot  \overrightarrow{\sigma_1}
+ b \overrightarrow{\beta_2} \cdot  \overrightarrow{\sigma} + 
ab (\overrightarrow{\alpha} \cdot  \overrightarrow{\sigma_1})
(\overrightarrow{\beta} \cdot  \overrightarrow{\sigma_2})
\rangle_{\psi_{-}} \nonumber \\
& = & \frac{1}{4}(1+ab\langle 
(\overrightarrow{\alpha} \cdot  \overrightarrow{\sigma_1})
(\overrightarrow{\beta} \cdot  \overrightarrow{\sigma_2})
\rangle_{\psi_{-}}) = 
\frac{1}{4}(1+ab\langle AB \rangle)
\end{eqnarray} ($\langle 
\overrightarrow{(\alpha} \cdot  \overrightarrow{\sigma_1})
(\overrightarrow{\beta} \cdot  \overrightarrow{\sigma_2})
\rangle_{\psi_{-}} =
- \overrightarrow{\alpha} \cdot \overrightarrow{\beta}$).

To summarize, 

\begin{eqnarray}
P^{AB}_{++}=P^{AB}_{--}=\frac{1}{4}(1+\langle A_1B_2 \rangle) \nonumber \\
P^{AB}_{+-}=P^{AB}_{-+}=\frac{1}{4}(1-\langle A_1B_2 \rangle); \nonumber \\
P^{AC}_{++}=P^{AC}_{--}=\frac{1}{4}(1+\langle A_1C_2 \rangle) \nonumber \\
P^{AC}_{+-}=P^{AC}_{-+}=\frac{1}{4}(1-\langle A_1C_2 \rangle); \nonumber \\
P^{BC}_{++}=P^{BC}_{--}=\frac{1}{4}(1-\langle B_1C_2 \rangle)  \nonumber \\
P^{BC}_{+-}=P^{BC}_{-+}=\frac{1}{4}(1+\langle B_1C_2 \rangle); 
\end{eqnarray}
Note the opposite signs in the last two lines. That is due to the fact that we had defined
$P^{BC} \equiv P^{B_2 C_2} = P^{(-B_1)C_2}$.

It is now straightforward to see that equations (\ref{consistency}) are satisfied.
We are thus assured of the existence of our quasiprobabilities.

Finally, the inequalities $8 P^{ABC}_{abc}= 8 (M^+ p+c x_h)_{abc} \geq 0$ ($a,b,c=+,-$), become:

\begin{eqnarray}
1+\langle AB \rangle+\langle AC \rangle-\langle BC \rangle-c \geq0 \nonumber \\
1+\langle AB \rangle-\langle AC \rangle+\langle BC\rangle+c\geq0 \nonumber \\
1-\langle AB \rangle+\langle AC \rangle+\langle BC\rangle+c\geq0 \nonumber \\
1-\langle AB \rangle-\langle AC \rangle-\langle BC\rangle-c\geq0 \nonumber \\
1-\langle AB \rangle-\langle AC \rangle-\langle BC\rangle+c\geq0 \nonumber \\
1-\langle AB \rangle+\langle AC \rangle+\langle BC\rangle-c\geq0 \nonumber \\
1+\langle AB \rangle-\langle AC \rangle+\langle BC\rangle-c\geq0 \nonumber \\
1+\langle AB \rangle+\langle AC \rangle-\langle BC\rangle+c\geq0 \label{ielist}
\end{eqnarray}The question: is there any value of $c$ such that all these inequalities are
simultaneously satisfied.

The first and last equations imply:
\begin{equation}
1 + \langle AB \rangle \geq  -(\langle AC \rangle - \langle BC\rangle) 
\end{equation}Similarly, the second and second to last inequalities imply:
\begin{equation}
1 + \langle AB \rangle \geq  +(\langle AC \rangle - \langle BC\rangle). 
\end{equation} Hence, together they are nothing other than Bell's famous inequality:

\begin{equation}
1 + \langle AB \rangle \geq  |\langle AC \rangle - \langle BC\rangle|. \label{BellIE} 
\end{equation}The remaining four inequalities imply:

\begin{equation}
1 - \langle AB \rangle \geq  |\langle AC \rangle + \langle BC\rangle| 
\end{equation}which is inequality (\ref{BellIE}) with the substitution $B \mapsto -B$.

Conversely, the last two inequalities imply that inequalities (\ref{ielist}) are satisfied for $c=0$.

\section{Conclusion}

Because of the symmetries of the singlet state of two spin-1/2 particles, and
the existence of just two measurement results for each single variable - 
the joint distributions of two variables $\{A,B\}$ can be expressed in terms
of a single parameter, the \emph{linear} correlation $\langle AB\rangle$.
Therefore, the most general inequalities on the distributions of the pairs
$\{A,B\}, \{A,C\},\{B,C\}$ for them to be generated as marginals of a 
(hypothetical) joint distribution of $\{A,B,C\}$ can be stated in terms of 
$\langle AB \rangle, \langle AC \rangle$ and $ \langle BC \rangle$.
These are Bell's inequalities. However, the underlying assumptions behind the
derivation are much more obvious when stated in terms of the distributions 
rather than the correlations. Furthermore, there is a subtle psychological
danger of confusing the general question of correlations (i.e. any statistical
dependence) with the much more restricted sense of 'linear correlations'. 
Once observables with more than two eigenvalues are considered, these two 
concepts become quite distinct.

\begin{center}

ACKNOWLEDGMENTS

\vspace{.1 in}

It is my pleasure to thank Itamar Pitowsky for his helpful comments and for 
referring me to earlier works of this nature.
I gratefully acknowledge the support of the Robert A. Welch Foundation.

\end{center}

\end{document}